\documentclass{article}
\usepackage{graphics}
\usepackage{setspace}
\title{ Interaction of Fermionic Matter and ECSK Black Hole with Torsion }
\author{ Emre Dil \\ Faculty of Engineering, Beykent University, \\34398, Sariyer, Istanbul-TURKEY \\ e-mail: emredil@beykent.edu.tr  \\[2ex]
         }

\begin{document}

\maketitle
\begin{abstract}
The interactions between the spin of fermionic matter and torsion in the Einstein--Cartan--Sciama--Kibble (ECSK) theory of gravity provides a repulsive gravitational potential at the very dense states of fermionic matter, which prevents the formation of black hole singularities inside the deeper horizon. While the fermionic matter in the black hole is attracted by the black hole at the beginning, after a critical point it is repelled to bounce at a critical high density and then expand into other side of the horizon as a newly created space, which may be considered as a nonsingular, closed universe. We constructed the action of these fermions in a black hole with torsion in the framework of ECSK theory of gravity from which the free Dirac action is inferred to obtain the interaction potential. The creation of a bouncing universe with the extremely repulsive potential may be related to the running vacuum and the modified Friedmann equations yield the consistent cosmological parameters with present FRW universe. Finally, this scenario naturally solves the flatness and horizon problems of cosmology without introducing finely tuned scalar fields, or more complicated functions of the Ricci Scalar R in the gravitational action.
\end{abstract}


\section{Introduction}

In curved space-time, the conservation law for the total angular momentum of fermions represented by the Dirac equation requires an asymmetric affine connection. Because metric general relativity has constraints on the symmetry of the connection, the ECSK theory of gravity correspondingly fixes the constraint of general relativity. ECKS relates the spin density of fermionic matter and the antisymmetric part of the affine connection as the torsion tensor, and assigns it as a dynamical variable like the metric itself. The intrinsic spin of fermions generates a torsion for the space-time by coupling to the connection of the curved space-time. On the other hand, The ECSK theory and general relativity give same predictions for the high densities as in the nuclear scales due to the negligibly small contribution from torsion to the Einstein equations, therefore the ECKS is validated for all cases as is general relativity [1-19].

 If the densities are extremely high as in the black holes and as in the very early universe, this situation leads to a spinor--torsion interaction behaves like a gravitational repulsion, and it prevents the creation of singularities in the black holes due to the spin of the fermionic matter [20-31]. It leads to that the singular big bang is turned out to be a nonsingular big bounce, before which the universe was contracting [32-35]. This model also provides a solution for the flatness-horizon problem of cosmology other than terminating the initial singularity of the universe [29-31], [34-37]. Instead of the cosmic inflation which is verified by CMB radiation inhomogeneities and requires additional scalar field matter components for solving the flatness-horizon problem [38-40], torsion can be considered as the simplest and most natural mechanism to provide solution for these major problems of the standard big bang cosmology [17,41,42].

 There can be found noteworthy studies which consider the nonsingular and bouncing cosmologies led by a universe inside a black hole whose event horizon opens to another new universe as a white hole [43-46]. The contraction of a universe in a black hole could correspond to gravitational collapse of matter inside the black hole existing in another universe. Therefore, the collapsing matter inside the black hole should have a bounce at a finite density, after that should expand into a new region on the other side of the event horizon, which can be considered as a nonsingular, new born closed universe [17,26,32,35], [47-54]. This is why the ECSK theory is expected to give the evolution of each spatial point in the universe toward a state of an extremely high but finite density due to the spinor-torsion interaction of fermions in the black hole. For such a scenario, the local contraction of universe finished, and the matter evolves toward a bounce, and the expansion of universe starts.

 In this paper we will be concerned with the behavior of fermion fields near the horizons of a spherically symmetric space-time with torsion in the generalized Einstein-Cartan-Kibble-Sciama theory of gravity by considering higher order terms to investigate the big bounce behavior of the universe. We will study the properties of a spinor field equation in the background of an ECSK black hole by considering the spin-torsion interactions of the conventional fermion field and the space-time.

 \section{Spinor Field in ECSK formalism}

 In order to study the dynamics of spinor fields in curved space-times we should use the formalism of ECSK theory, since it holds for the sources with spin. In the curved space-times the theory of Dirac spinors is a complex example of the quantum field theories for obtaining the energy-momentum tensor of the field from the variation of spinor Lagrangian. Although the energy-momentum tensor of scalar fields describes the reaction of Lagrangian to the variations of the metric, as the field is itself held constant during the variation of the metric, the same does not apply for the spinor field energy-momentum tensor from the variation of metric tensor only, where the spinor fields are the sections of a spinor bundle. This spinor bundle is obtained as an associated vector bundle due to the bundle of tetrad spin frames. The bundle of spin frames is a double covering of the bundle of oriented and time-oriented orthonormal frames. For spinor fields, when one varies the metric, the components of the spinor fields also varies and they cannot be held fixed with respect to some fixed holonomic frame as in the scalar field case [55]. Therefore, we need to give the algebraic structure of the ECSK formalism and the spinor field structures.

 The metric-affine formulation of the gravity has the dynamical variables of the tetrad frame field $e_{a}^{i} $ and the spin connection $\omega _{bk}^{a} =e_{j}^{a} e_{b;k}^{j} =e_{j}^{a} (e_{b,k}^{j} +\Gamma _{ik}^{j} e_{b}^{i} )$, where the comma denotes the ordinary partial derivative with respect to the $x^{k} $ coordinate, and the semicolon represents the covariant derivative with respect to the affine connection $\Gamma _{jk}^{i} $ whose antisymmetric lower indices give the torsion tensor $S_{jk}^{i} =\Gamma _{[jk]}^{i} $. Moreover, the tetrad frames link the space-time coordinates with indices $i,j,...$ and the local Lorentz coordinates with indices $a,b,...$, such that $V^{a} =V^{i} e_{i}^{a} $ for a Lorentz vector $V^{a} $ and a standard vector $V^{i} $. While the Lorentz vectors have a covariant derivative denoted by a bar and defined in terms of the spin connection: $V_{\left|i\right. }^{a} =V_{,i}^{a} +\omega _{bi}^{a} \, V^{b} $ and $V_{a\left|i\right. } =V_{a,i} -\omega _{ai}^{b} \, V_{b} $, the standard vectors have a semicolon covariant derivative defined in terms of the affine connection, $V_{;i}^{k} =V_{,i}^{k} +\Gamma _{li}^{k} \, V^{l} $ and $V_{k;i} =V_{k,i} -\Gamma _{ki}^{l} \, V_{l} $. The Local Lorentz coordinates are lowered-raised by the Minkowski metric $\eta _{ab} $, and the space-time coordinates are lowered-raised by the metric tensor $g_{ik} $. Also, the metricity condition $g_{i\, j;k} =0$ yields the affine connection defined to be $\Gamma _{i\, j}^{k} =\{ _{i\, j}^{k} \} +C_{i\, j}^{k} $ in terms of the Christoffel symbols $\{ _{i\, j}^{k} \} =(1/2)g^{km} (g_{mi,j} +g_{m\, j,i} -g_{i\, j,m} )$ and the contortion tensor $C_{jk}^{i} =S_{jk}^{i} +2S_{(jk)}^{i} $. While the symmetrization is used by $A_{(jk)} =(1/2)(A_{jk} +A_{k\, j} )$, the antisymmetrization is used by $A_{[jk]} =(1/2)(A_{jk} -A_{k\, j} )$throughout the paper. In ECSK gravity, the metric $g_{ik} =\eta _{ab} e_{i}^{a} e_{k}^{b} $ and the torsion $S_{ik}^{j} =\omega _{[ik]}^{j} +e_{[i,k]}^{a} e_{a}^{j} $ can also be taken as the dynamical variables more than the tetrad frames and the spin connection [1-19].

 From the definition, a tensor density $T_{D} $ is related to the corresponding tensor $T$ by $T_{D} =eT$, where $e=\det e_{i}^{a} =\sqrt{-\det g_{ik} } $, then the spin density and the energy-momentum densities are given as $\sigma _{i\, jk} =e\, s_{i\, jk} $ and ${\rm T} _{ik} =eT_{ik} $. These tensors are called as \textit{metric} spin tensor and \textit{metric} energy-momentum tensor since they are specified by the space-time coordinate indices, and they are obtained from the variation of the Lagrangian with respect to the torsion (or contortion) tensor $C_{k}^{i\, j} $ and the metric tensor $g^{i\, j} $, respectively. Accordingly, the metric spin tensor is given as $s_{i\, j}^{k} =(2/e)(\delta \ell _{m} /\delta C_{k}^{i\, j} )=(2/e)(\partial \ell _{m} /\partial C_{k}^{i\, j} )$, while the metric energy-momentum tensor is $T_{i\, j} =(2/e)(\delta \ell _{m} /\delta g^{i\, j} )=(2/e)[\partial \ell _{m} /\partial g^{i\, j} -\partial _{k} (\partial \ell _{m} /\partial (g_{,k}^{i\, j} ))]$. The Lagrangian density of the source matter field is here $\ell _{m} =eL_{m} $. If the local Lorentz coordinates are used to specify these tensors as $\sigma _{ab}^{i} =e\, s_{ab}^{i} $ and ${\rm T} _{i}^{a} =eT_{i}^{a} $, then $s_{ab}^{i} $ and $T_{i}^{a} $ are called as the \textit{dynamical} spin tensor and \textit{dynamical} energy-momentum tensor, respectively, and they are obtained from the variation of the Lagrangian with respect to the tetrad $e_{a}^{i} $ and the spin connection $\omega _{i}^{ab} $, such that $s_{ab}^{i\, } =(2/e)(\delta \ell _{m} /\delta \omega _{i}^{ab} )=(2/e)(\partial \ell _{m} /\partial \omega _{i}^{ab} )$, and $T_{i}^{a} =(1/e)(\delta \ell _{m} /\delta e_{a}^{i} )=(1/e)[\partial \ell _{m} /\partial e_{a}^{i} -\partial _{j} (\partial \ell _{m} /\partial (e_{a,j}^{i} ))]$ [1-19].

 Total action of the gravitational field with spinor field dark matter in metric-affine ECSK theory is given in the same form with the classical Einstein-Hilbert action, such as $S=\kappa \int (\ell _{g} +\ell _{\psi } )d^{4} x $, where $\kappa =8\pi \, G$ and $\ell _{g} =-(1/2\kappa )eR$, and $\ell _{\psi } $ are the gravitational, and fermionic matter Lagrangian densities. The Ricci scalar is $R=R_{j}^{b} e_{b}^{j} $ where $R_{j}^{b} =R_{jk}^{bc} e_{c}^{k} $ is the Ricci tensor obtained from the curvature tensor $R_{jk}^{bc} $. Also, the curvature tensor is related to the spin connection, such that $R_{bi\, j}^{a} =\omega _{bj,i}^{a} -\omega _{bi,j}^{a} +\omega _{ci}^{a} \, \omega _{bj}^{c} -\omega _{cj}^{a} \, \omega _{bi}^{c} $. The variation of the action with respect to the contortion tensor gives the Cartan equations $S_{i\, k}^{j} -S_{i} \, \delta _{k}^{j} +S_{k} \, \delta _{i}^{j} =-(\kappa /2e)\, \sigma _{ik}^{j} $, and the variation with respect to the metric tensor yields the Einstein equations $G_{ik} =\kappa (T_{ik}^{\psi } +U_{ik}^{\psi } )$, where $G_{ik} =P_{i\, jk}^{j} -(1/2)P_{lm}^{lm} g_{ik} $ is the Einstein tensor. Here $P_{i\, jk}^{j} $ is also the Riemann curvature tensor given by the relation $R_{klm}^{i} =P_{klm}^{i} +C_{km\, :l}^{i} -C_{kl\, :\, m}^{i} +C_{km}^{j} C_{jl}^{i} -C_{kl}^{j} C_{jm}^{i} $, where colon represents the Riemannian covariant derivative with respect to the Levi-Civita connection $\{ _{li}^{k} \} \, $, such as $V_{:i}^{k} =V_{,i}^{k} +\{ _{li}^{k} \} \, V^{l} $ and $V_{k\, :i} =V_{k,i} -\{ _{ki}^{l} \} \, V_{l} $. Also, the curvature tensor transforms into the Riemann tensor for torsion-free general relativity theory. The $U_{ik} $ term in the Einstein equations is the spin contribution $U_{ik} =\kappa (-s_{[l}^{ij} s_{j]}^{kl} -(1/2)s^{i\, jl} s_{jl}^{k} +(1/4)s^{jli} s_{jl}^{k} +(1/8)g^{ik} (-4s_{j[m}^{l} s_{l]}^{jm} +s^{jlm} s_{jlm} )$, and the total energy-momentum tensor of the spinor field is given by $\Theta _{ik}^{\psi } =T_{ik}^{\psi } +U_{ik}^{\psi } $ [1-19].

 Combining the above algebraic relations for the metric-affine ECSK formulation of gravity for $(+,-,-,-)$ metric signature, spinor field fermionic matter is described by the Lagrangian densities of the form:
\begin{equation} \label{GrindEQ__1_} 
\ell _{\psi } =e(i/2)(\bar{\psi }\gamma ^{k} \psi _{;k} -\bar{\psi }_{;k} \gamma ^{k} \psi )-em_{\psi } \bar{\psi }\psi ,     
\end{equation} 
where $\psi $ and $\bar{\psi }=\psi ^{+} \gamma ^{0} $ are the spinor and the adjoint spinor fields, respectively. The semicolon covariant derivative of the spinor and the adjoint spinor fields are given by 
\begin{equation} \label{GrindEQ__2_} 
\psi _{;\, k} =\psi _{,\, k} -\Gamma _{k} \psi ,      
\end{equation} 
\begin{equation} \label{GrindEQ__3_} 
\bar{\psi }_{;\, k} =\bar{\psi }_{,\, k} -\Gamma _{k} \bar{\psi },      
\end{equation} 
where $\Gamma _{k} =-(1/4)\, \omega _{abk} \gamma ^{a} \gamma ^{b} $ is the Fock-Ivanenko spin connection, $\gamma ^{k} $ and $\gamma ^{a} $ are the \textit{metric} and \textit{dynamical} Dirac gamma matrices as; $\gamma ^{k} =e_{a}^{k} \gamma ^{a} $, $\gamma ^{(k} \gamma ^{m)} =g^{km} I$ and $\gamma ^{(a} \gamma ^{b)} =\eta ^{ab} I$. One can decompose the semicolon covariant derivative of the spinor field into a colon Riemannian covariant derivative with the contortion tensor $C_{i\, jk} $ term as
\begin{equation} \label{GrindEQ__4_} 
\psi _{;\, k} =\psi _{:\, k} +(1/4)C_{i\, jk} \gamma ^{[i} \gamma ^{j]} \psi ,           
\end{equation} 
\begin{equation} \label{GrindEQ__5_} 
\bar{\psi }_{;\, k} =\bar{\psi }_{:\, k} -(1/4)C_{i\, jk} \bar{\psi }\gamma ^{[i} \gamma ^{j]} .           
\end{equation} 
The colon Riemannian covariant derivative is also defined to be 
\begin{equation} \label{GrindEQ__6_} 
\psi _{:\, k} =\psi _{,\, k} +(1/4)\, g_{ik} \{ _{jm}^{i} \} \gamma ^{j} \gamma ^{m} \psi ,             
\end{equation} 
\begin{equation} \label{GrindEQ__7_} 
\bar{\psi }_{:\, k} =\bar{\psi }_{,\, k} -(1/4)\, g_{ik} \{ _{jm}^{i} \} \gamma ^{j} \gamma ^{m} \bar{\psi }.             
\end{equation} 
Although the spinor field Lagrange density contains covariant derivatives including the contortion tensor $C_{i\, jk} $, the explicit form of the contortion tensor is obtained from the Cartan equations whose right hand side involves the spin tensor density. Then, the spin tensor is led by the variation of the spinor Lagrangian with respect to the contortion tensor, such as
\begin{equation} \label{GrindEQ__8_} 
s^{i\, jk} =(1/e)\sigma ^{i\, \, jk} =-(1/e)\, \varepsilon ^{i\, jkl} s_{l} ,            
\end{equation} 
where $\varepsilon ^{i\, jkl} $ is the Levi-Civita symbol, and
\begin{equation} \label{GrindEQ__9_} 
s^{i} =(1/2)\, \bar{\psi }\gamma ^{i} \gamma ^{5} \psi  
\end{equation} 
is the spin pseudo-vector, and $\gamma ^{5} =i\gamma ^{0} \gamma ^{1} \gamma ^{2} \gamma ^{3} $. Substituting the spin tensor of spinor field in the Cartan equations leads to the torsion tensor as
\begin{equation} \label{GrindEQ__10_} 
S_{i\, jk} =C_{i\, jk} =(1/2)\kappa \varepsilon _{i\, jkl} s^{l} ,     
\end{equation} 
which will be found in the spinor field Lagrange density [1-19].

Then, the variation of the spinor fermionic matter Lagrangian density with respect to the adjoint spinor $(\partial \ell _{\psi } /\partial \bar{\psi })-(\partial \ell _{\psi } /\partial \bar{\psi }_{:\, k} )_{:\, k} =0$ gives the ECSK Dirac equation
\begin{equation} \label{GrindEQ__11_} 
i\gamma ^{k} \psi _{:k} -m_{\psi } \psi +\frac{3}{8} \kappa (\bar{\psi }\gamma ^{k} \gamma ^{5} \psi )\gamma _{k} \gamma ^{5} \psi =0,    
\end{equation} 
while the variation with respect to the spinor itself $(\partial \ell _{\psi } /\partial \psi )-(\partial \ell _{\psi } /\partial \psi _{:\, k} )_{:\, k} =0$ gives adjoint ECSK Dirac equation as
\begin{equation} \label{GrindEQ__12_} 
i\bar{\psi }_{:k} \gamma ^{k} +m_{\psi } \bar{\psi }-\frac{3}{8} \kappa (\bar{\psi }\gamma ^{k} \gamma ^{5} \psi )\, \bar{\psi }\gamma _{k} \gamma ^{5} =0.    
\end{equation}

 \section{Space-time of an ECSK Black Hole with Torsion}

 A closed universe with quantum effects of fermionic spinor field in a curved space-time of a black hole provides an oscillatory universe as the big bounce. After the big bounce a first accelerated expansion phase of the universe becomes in a torsion-dominated era. The fermions composing the spin fluid prevents to form a singularity in the black hole due to the spin-torsion coupling. Moreover, the spin--torsion term triggers a big bounce from the other side of the event horizon after the universe collapse into minimum scale in the black hole [17,41].

Therefore, it is necessary to define a black hole with torsion in the ECSK formalism which corresponds an action [56-60]
\begin{equation} \label{GrindEQ__13_} 
S_{BH} =\int e\, \left[-\frac{1}{2\kappa } \left(L+a_{1} L_{1} \right)\right]\, d^{4} x ,            
\end{equation} 
where $a_{1} $ is a coupling constant, $L$ is ECSK Lagrangian and it is given as
\begin{equation} \label{GrindEQ__14_} 
L=R+\frac{1}{4} S_{i\, jk} S^{i\, jk} +\frac{1}{2} S_{i\, jk} S^{ji\, k} +S_{i}^{i\, j} S_{jk}^{k} +2S_{i;j}^{i\, j} ,           
\end{equation} 
and $L_{1} $ is
\begin{equation} \label{GrindEQ__15_} 
L_{1} =RS_{i\, jk} S^{i\, jk} +\frac{1}{4} S_{i\, jk} S^{i\, jk} \left[S^{l\, mn} \left(2S_{m\ln } +S_{l\, mn} \right)+8S_{l;m}^{l\, m} \right]+S_{i}^{i\, j} S_{jk}^{k} S^{l\, mn} S_{l\, mn} .      
\end{equation} 
The action of an ECSK black hole yields a static spherically symmetric space-time with the metric
\begin{equation} \label{GrindEQ__16_} 
d\, s^{2} =H(r)\, dt^{2} -\frac{dr^{2} }{F(r)} -r^{2} \left(d\theta ^{2} +\sin ^{2} \theta \, d\phi ^{2} \right),           
\end{equation} 
where  [60]
\[F(r)=1-\frac{c_{1} }{c_{2} \sqrt{r} } +\frac{6c_{3} c_{2}^{2} -9c_{1}^{2} \ln (3c_{1} -2c_{2} \sqrt{r} )}{6c_{2}^{2} r} ,\] 
\begin{equation} \label{GrindEQ__17_} 
H(r)=\left(1-\frac{2c_{1} }{c_{2} \sqrt{r} } \right)^{2} .               
\end{equation} 
Here $c_{1} $, $c_{2} $ and $c_{3} $ are the integration constants related to the torsion. There exists a constraint on these constants due to the logarithmic term in $F(r)$, such that, the term $3c_{1} -2c_{2} \sqrt{r} $ must be positive at the physical region. If $c_{2} >0$ is assumed for this term, then $c_{1} <0$ always gives negative $3c_{1} -2c_{2} \sqrt{r} $, but $c_{1} >0$ may give a positive $3c_{1} -2c_{2} \sqrt{r} $ with the constraint $\sqrt{r} <3c_{1} /2c_{2} $ yielding a divergent $F(r)$ function. Because of the divergent nature of $F(r)$, $c_{2} <0$ limit is considered instead of $c_{2} >0$ limit. In addition, the coupling constant $a_{1} $ in \ref{GrindEQ__13_} is taken as $a_{1} =2/9c_{2}^{2} $ in order for the metric \ref{GrindEQ__16_} to be asymptotically flat at the spatial infinity for $c_{2} <0$ limit.

 It is assumed that the $c_{2} <0$ limit refers to $c_{2} =-1$ throughout the paper, for convenience. For the event horizon of the black hole, the $F(r)=0$ case is taken into account and the critical values for constant $c_{3} $ are obtained from the numerical analysis because of the complexity of the logarithmic term:
\begin{equation} \label{GrindEQ__18_} 
c_{crit} =\left\{\begin{array}{c} {\frac{3c_{1}^{2} }{2} \ln (3c_{1} ),\qquad c_{1} >0} \\ {\frac{c_{1}^{2} }{2} \left[3\ln (-c_{1} )-4\right],\qquad c_{1} <0} \end{array}\right. .       
\end{equation} 
For the case $c_{3} >c_{crit} $, no horizon exists in the space-time and the geometry is a naked singularity. However, for the case $c_{3} <c_{crit} $ the black hole has one horizon in $c_{1} >0$ limit, and two horizons in $c_{1} <0$ limit [56-60].

 \section{Spinor Field in an ECSK Black Hole Background}

 Bouncing universe is proposed to be led by the existence of spinor matter in an ECSK black hole with torsion. In order to investigate the validity of our assumption, we now consider the spinor field \ref{GrindEQ__1_} in the ECSK black hole background \ref{GrindEQ__16_} with the action
\begin{equation} \label{GrindEQ__19_} 
S=\int e\, \left(\frac{i}{2} (\bar{\psi }\gamma ^{k} \psi _{;k} -\bar{\psi }_{;k} \gamma ^{k} \psi )-m_{\psi } \bar{\psi }\psi \right)d^{4} x ,           
\end{equation} 
where
\begin{equation} \label{GrindEQ__20_} 
e=\sqrt{\frac{H(r)\, }{F(r)} } r^{2} \sin \theta ,      
\end{equation} 
for the metric of ECSK black hole. We expand the semicolon covariant derivatives in the action \ref{GrindEQ__19_} by using the equations \ref{GrindEQ__2_}-\ref{GrindEQ__10_}, such that
\begin{equation} \label{GrindEQ__21_} 
\psi _{;\, k} =\psi _{,\, k} +\frac{1}{4} \left[\, g_{ik} \{ _{jm}^{i} \} \gamma ^{j} \gamma ^{m} +\frac{1}{4} \kappa \varepsilon _{ijkl} \bar{\psi }\gamma ^{l} \gamma ^{5} \psi \, \gamma ^{\left[i\right. } \gamma ^{\left. j\right]} \right]\, \psi ,           
\end{equation} 
\begin{equation} \label{GrindEQ__22_} 
\bar{\psi }_{;\, k} =\bar{\psi }_{,\, k} -\bar{\psi }\frac{1}{4} \left[\, g_{ik} \{ _{jm}^{i} \} \gamma ^{j} \gamma ^{m} +\frac{1}{4} \kappa \varepsilon _{ijkl} \bar{\psi }\gamma ^{l} \gamma ^{5} \psi \, \gamma ^{\left[i\right. } \gamma ^{\left. j\right]} \right].           
\end{equation} 
We then take the derivatives for $k=0,1,2,3$ components in \ref{GrindEQ__19_} with \ref{GrindEQ__21_} and \ref{GrindEQ__23_}. For the metric \ref{GrindEQ__16_}, $\{ _{jm}^{i} \} \, $Christoffel connections are obtained as
\[\{ _{01}^{0} \} =\{ _{10}^{0} \} =\frac{1}{2} \frac{H'(r)}{H(r)} ,\] 
\[\{ _{00}^{1} \} =\frac{1}{2} H'(r)\, F(r), \{ _{11}^{1} \} =-\frac{1}{2} \frac{F'(r)}{F(r)} , \{ _{22}^{1} \} =-r\, F(r), \{ _{33}^{1} \} =-r\, F(r)\sin ^{2} \theta ,\] 
\[\{ _{12}^{2} \} =\{ _{21}^{2} \} =\{ _{13}^{3} \} =\{ _{31}^{3} \} =\frac{1}{r} ,\] 
\[\{ _{33}^{2} \} =-\sin \theta \cos \theta ,\] 
\begin{equation} \label{GrindEQ__23_} 
\{ _{23}^{3} \} =\{ _{32}^{3} \} =\cot \theta .       
\end{equation} 
By applying the identities about the components of $\gamma $ matrices given below equation \ref{GrindEQ__3_} and \ref{GrindEQ__9_}, we find the components of semicolon covariant derivatives of spinor field such that
\begin{equation} \label{GrindEQ__24_} 
\psi _{;0} =\psi _{,0} ,            
\end{equation} 
\begin{equation} \label{GrindEQ__25_} 
\bar{\psi }_{;0} =\bar{\psi }_{,0} ,            
\end{equation} 
\begin{equation} \label{GrindEQ__26_} 
\psi _{;1} =\psi _{,1} -\left[\frac{H'(r)}{8H(r)} +\frac{F'(r)}{8F(r)} +\frac{1}{2r} \right]\psi ,        
\end{equation} 
\begin{equation} \label{GrindEQ__27_} 
\bar{\psi }_{;1} =\bar{\psi }_{,1} +\bar{\psi }\left[\frac{H'(r)}{8H(r)} +\frac{F'(r)}{8F(r)} +\frac{1}{2r} \right],       
\end{equation} 
\begin{equation} \label{GrindEQ__28_} 
\psi _{;2} =\psi _{,2} +\frac{1}{4} \frac{\cot \theta }{r^{2} } \psi ,        
\end{equation} 
\begin{equation} \label{GrindEQ__29_} 
\bar{\psi }_{;2} =\bar{\psi }_{,2} -\frac{1}{4} \bar{\psi }\frac{\cot \theta }{r^{2} } ,        
\end{equation} 
\begin{equation} \label{GrindEQ__30_} 
\psi _{;3} =\psi _{,3} ,            
\end{equation} 
\begin{equation} \label{GrindEQ__31_} 
\bar{\psi }_{;3} =\bar{\psi }_{,3} .            
\end{equation} 
We now substitute these semicolon covariant derivative components into the action \ref{GrindEQ__19_}, then
\begin{equation} \label{GrindEQ__32_} 
\begin{array}{l} {S=\int \, \left(\frac{i}{2} (\bar{\psi }\gamma ^{0} \psi _{,0} -\bar{\psi }_{,0} \gamma ^{0} \psi +\bar{\psi }\gamma ^{1} \psi _{,1} -\bar{\psi }_{,1} \gamma ^{1} \psi -\left[\frac{H'(r)}{4H(r)} +\frac{F'(r)}{4F(r)} +\frac{1}{r} \right]\bar{\psi }\gamma ^{1} \psi \right.  } \\ {\, \, \, \, \, \, \, \, \, \, \, \, \, \, \, \, \, \, \, \, \left. +\bar{\psi }\gamma ^{2} \psi _{,2} -\bar{\psi }_{,2} \gamma ^{2} \psi +\frac{1}{2} \frac{\cot \theta }{r^{2} } \bar{\psi }\gamma ^{2} \psi +\bar{\psi }\gamma ^{3} \psi _{,3} -\bar{\psi }_{,3} \gamma ^{3} \psi )-m_{\psi } \bar{\psi }\psi \right)d^{4} x} \end{array},  
\end{equation} 
and
\begin{equation} \label{GrindEQ__33_} 
S=\int \, \left(\frac{i}{2} (\bar{\psi }\gamma ^{k} \psi _{,k} -\bar{\psi }_{,k} \gamma ^{k} \psi )-m_{\psi } \bar{\psi }\psi +V_{k} \bar{\psi }\gamma ^{k} \psi \right)d^{4} x .           
\end{equation} 
where the free action is inferred from \ref{GrindEQ__32_} as
\begin{equation} \label{GrindEQ__34_} 
S_{free} =\int \, \left(\frac{i}{2} (\bar{\psi }\gamma ^{k} \psi _{,k} -\bar{\psi }_{,k} \gamma ^{k} \psi )-m_{\psi } \bar{\psi }\psi \right)d^{4} x .              
\end{equation} 
Therefore, the remaining terms in \ref{GrindEQ__32_} gives the vector potential of the spinor matter due to an ECSK black hole geometry with the components
\begin{equation} \label{GrindEQ__35_} 
V_{r} =-\left[\frac{H'(r)}{4H(r)} +\frac{F'(r)}{4F(r)} +\frac{1}{r} \right],  V_{\theta } =\frac{1}{2} \frac{\cot \theta }{r^{2} } .       
\end{equation} 
For the case $c_{2} =-1$ and $c_{1} >0$ one-horizon case with the use of \ref{GrindEQ__18_}, one can obtain $F(r)$, $H(r)$ and derivatives, as
\[F(r)=1+\frac{c_{1} }{\sqrt{r} } +\frac{9c_{1}^{2} \ln (3c_{1} )-9c_{1}^{2} \ln (3c_{1} +2\sqrt{r} )}{6r} ,\] 
\begin{equation} \label{GrindEQ__36_} 
H(r)=\left(1+\frac{2c_{1} }{\sqrt{r} } \right)^{2} ,               
\end{equation} 
and
\[F'(r)=-\frac{c_{1} }{2r^{3/2} } -\frac{3c_{1}^{2} }{6c_{1} r^{3/2} +4r^{4/2} } -\frac{3c_{1}^{2} \ln (3c_{1} )}{2r^{4/2} } +\frac{3c_{1}^{2} \ln (3c_{1} +\sqrt{r} )}{2r^{4/2} } ,\] 
\begin{equation} \label{GrindEQ__37_} 
H'(r)=-\frac{2c_{1} }{r^{3/2} } -\frac{4c_{1}^{2} }{r^{4/2} } .               
\end{equation} 
Using \ref{GrindEQ__36_} and \ref{GrindEQ__37_} there can be found the components of the vector potential in terms of the torsion related constants. For the radial component of the potential \ref{GrindEQ__35_} the behavior of the potential with respect to radius is illustrated in Figure 1 for various positive $c_{1} $ values. As can be obtained from \ref{GrindEQ__18_} these constants lead to the Schwarzschild radius of the ECSK black hole of $0.6c_{1} $, and the potential has minimum values at approximately for the $0.1c_{1} $, which are higher density states in deeper horizon. These minimum values occur for $c_{1} =2GM$ at $r=0.12GM$, for $c_{1} =2.5GM$ at $r=0.19GM$, for $c_{1} =3.0GM$ at $r=0.28GM$ and for $c_{1} =3.5GM$ at $r=0.38GM$. The potential values before $r=0.1c_{1} $ is extremely repulsive. This implies a gravitational repulsion inside of the horizon of the ECSK black hole after a critical value for the Schwarzschild radius. On the other hand, the tangential component of the potential gives also the similar behavior for $r\approx 0.1c_{1} $ but with the polar angle $\theta \approx 0.1\pi $.

\begin{figure}
\includegraphics{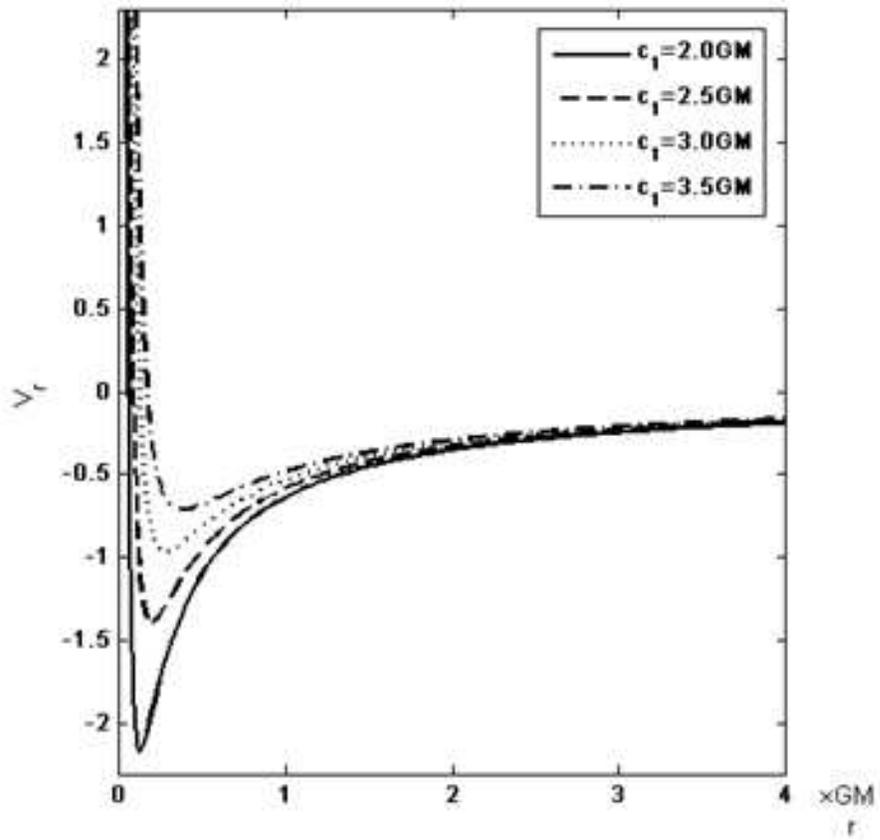}
\caption{Repulsive and Attractive Behavior of Radial Potential $V_{r} $}
\label{fig1}
\end{figure}

 This behavior expresses for the extremely high densities of fermions in the torsion dominated black holes as in the very early universe there leads to a spinor--torsion interaction behaves like a very strong gravitational repulsion which may prevent the creation of singularities in the black holes and cause to that the singular big bang is turned out to be a nonsingular big bounce from the other side of the horizon of the ECSK black hole.

 A closed universe in an ECSK black hole with particle production identifies an expanding universe led by the quantum effects in curved space-time. The extreme gravitational repulsion due to the high density fermions in the black hole exhausts from the other side of the horizon as fermion production [61-66]. The initial expansion of the closed universe defines the torsion-dominated era and the spin-torsion term results from the spin density of the fermionic spinor field which is given as
\begin{equation} \label{GrindEQ__38_} 
s^{2} =\frac{1}{8} (\hbar cn_{f} )^{2} ,               
\end{equation} 
where $n_{f} $ is the number density of fermions and the effective energy density, and also pressure of the spinor fluid read
\begin{equation} \label{GrindEQ__39_} 
\tilde{\varepsilon }=\varepsilon -\alpha \, n_{f}^{2} ,  \tilde{p}=p-\alpha \, n_{f}^{2}  ,              
\end{equation} 
where $\alpha =\kappa \, (\hbar c)^{2} /32$ and $\alpha \, n_{f}^{2} $ is the spin-torsion coupling term. The spin-torsion term decreases faster than $\varepsilon $ during the expansion phase.  The universe enters the phase of acceleration to expand infinity with the repulsive potential $V_{r} $ which may be thought as it is turning out to be a running vacuum as the varying cosmological constant $\Lambda \, (r)$ which has been previously proposed by us and other studies in the literature [67-70]. Then, the other side of the horizon creates a bouncing FRW universe with an accelerated expansion obeying the usual Friedmann equation with the running vacuum $\Lambda \, (r)$,
\begin{equation} \label{GrindEQ__40_} 
\frac{\dot{a}^{2} }{c^{2} } +1=\frac{D}{a} +\frac{1}{3} \Lambda \, (r)a^{2} ,                  
\end{equation} 
where
\begin{equation} \label{GrindEQ__41_} 
D=\frac{1}{3} \kappa \, h_{*} T_{eq} \left(\frac{\tilde{a}_{i} }{a_{i} } \right)^{3} (a_{i} T_{i} )^{3}  
\end{equation} 
comes from the spinor matter energy density. In equation \ref{GrindEQ__41_}, $h_{*} =(\pi ^{2} /30)g_{*} k_{B}^{4} /(\hbar c)^{3} $ with $g_{*} =(7/8)g_{f} $, where $g_{f} =\sum _{i}\, g_{i}  $ is summed over fermions and $g_{i} $ is the spin states for each particle species $i$. Moreover, $\tilde{a}_{i} $ is the scale factor at temperature $T_{i} $ in the expanding phase. For the expanding phase $\tilde{a}_{i} $ is greater than the scale factor $a_{i} $ before the expansion begins. Then, $\tilde{a}_{i} >a_{i} $ represents the expansion of the universe with the fermion creation in the other side of the ECSK black hole horizon. The consistent resolutions of the cosmological parameters from \ref{GrindEQ__41_} is found in the references [17,26,34] as $T_{eq} =8820\, K$ and $\Lambda /\kappa =5.24\times 10^{-10} \, Pa$ which gives the ratio $\tilde{a}_{i} /a_{i} >10^{10} $ representing the expansion of the universe, and the deviation of the density parameter from the unity $\tilde{\Omega }_{\min } -1=(\Omega _{\min } -1)(\tilde{a}_{i} /a_{i} )^{2} <10^{-55} $ solving both flatness and horizon problems of cosmology.

\section{Conclusions}

 In this paper, we considered the ECKS theory of gravity with torsion and fermion field in a black hole. Using spin density of fermionic matter as the source of torsion leads to a natural physical interpretation which does not introduce additional fields or coupling constants to form a repulsive force causing accelerated expansion of universe. Therefore, we first give the dynamics of spinor field in ECSK theory of gravity as constructing action to give the matter Lagrangian and modified Dirac equation. After that, we explore the geometry of an ECSK black hole in which the spinor field couples to the torsion of the space time in order to produce a repulsive potential which will drive the newly created exhausted particles from the other side of the black hole horizon.

 With the completed tools in sections 2 and 3, we have investigated behavior of spinor field in the ECSK black hole which forms a bouncing universe with a radial potential $V_{r} $ whose behavior is illustrated in Figure 1. When the radial coordinate is less than the Schwarzschild radius by a factor of 0.1, the potential becomes extremely repulsive as the event horizon is approximately $0.6c_{1} $. For some particular values of $c_{1} =2GM$, $c_{1} =2.5GM$, $c_{1} =3.0GM$, and $c_{1} =3.5GM$ the repulsive potential forms at coordinates less than $r=0.12GM$, $r=0.19GM$, $r=0.28GM$ and $r=0.38GM$, respectively. At these high density regions in the black hole horizon strong repulsive potential behaves like a running vacuum of $\Lambda \, (r)$ launches the accelerated expansion of the bouncing universe from the back side of the black hole horizon.

 The modified Friedmann equations of this newly created bouncing universe provides the consistent cosmological parameters as scale factor and fixed flatness horizon problems with a values of $\tilde{a}_{i} /a_{i} >10^{10} $ and $\tilde{\Omega }_{\min } -1<10^{-55} $, respectively.

At extremely high densities as in the deeper horizon of the black holes with the fermionic matter, a significant gravitational repulsion is generated by spin and torsion interactions. This repulsive potential prevents to form the singularities in black holes or big bang of the universe. With the contribution of the gravitational repulsion exhausting fermions from the other side of the horizon create particles, and all black holes may construct a new bouncing universe on the other side of its event horizon [26-28]. In addition, it simultaneously solves the flatness and horizon problems without requiring finely tuned scalar fields, or modified version of the Ricci Scalar R in the action of gravitational field by more complex functions.

\eject
\textbf{\Large{References}}
 
 [1] Lord E. A., 1976, Tensors, Relativity and Cosmology, McGraw-Hill

 [2] Kibble T. W. B., 1961, J. Math. Phys. 2, 212

 [3] Sciama D. W., 1964, Rev. Mod. Phys. 36, 463

 [4] Sciama D. W., 1964, Rev. Mod. Phys. 36, 1103

 [5] Hehl F. W., 1971, Phys. Lett. A 36, 225

 [6] Hehl F. W., 1973, Gen. Relativ. Gravit. 4, 333

 [7] Hehl F. W., 1974, Gen. Relativ. Gravit. 5, 491

 [8] Hehl F. W., P. von der Heyde, Kerlick G. D., Nester J. M., 1976, Rev. 
 
 Mod. Phys. 48, 393.

 [9] Hehl F. W., Datta B. K., 1971, J. Math. Phys. 12, 1334

 [10] Sabbata V. de, Gasperini M., 1985, Introduction to Gravitation World 
 
 Scientific

 [11] Sabbata V. de, Sivaram C., 1994, Spin and Torsion in Gravitation, World 
 
 Scientific

 [12] Schouten J. A., Ricci-Calculus, 1954, Springer-Verlag

 [13] Shapiro I. L., 2002, Phys. Rep. 357,113

 [14] Hammond R. T., 2002, Rep. Prog. Phys. 65, 599

 [15] Sotiriou T. P., Faraoni V., 2010, Rev. Mod. Phys. 82, 451

 [16] Capozziello S., Laurentis M. De, 2011, Phys. Rept. 509, 167

 [17] Poplawski N. J., 2016, ApJ, 832, 96.

 [18] Poplawski, N. 2013, arXiv:1304.0047,

 [19] Poplawski, N. J. 2009, arXiv:0911.0334

 [20] Kopczy\'{n}ski, W. 1972, PhLA, 39, 219

 [21] Kopczy\'{n}ski, W. 1973, PhLA, 43, 63

 [22] Trautman, A. 1973, NPhS, 242, 7

 [23] Hehl, F. W. 1974, GReGr, 5, 491

 [24] Gasperini, M. 1986, PhRvL, 56, 2873

 [25] Brechet, S. D., Hobson, M. P., \& Lasenby, A. N. 2008, CQGra, 25, 
 
 245016

 [26] Poplawski, N. J. 2010, PhLB, 687, 110

 [27] Poplawski, N. 2012, PhRvD, 85, 107502

 [28] Poplawski, N. J. 2013, PhLB, 727, 575

 [29] Magueijo, J., Zlosnik, T. G., \& Kibble, T. W. B. 2013, PhRvD, 87, 
 
 063504

 [30] Alexander, S., Bambi, C., Marcian\`{o}, A., \& Modesto, L. 2014, PhRvD, 
 
 90, 123510

 [31] Bambi, C., Malafarina, D., Marcian\`{o}, A., \& Modesto, L. 2014, PhLB, 
 
 734, 27

 [32] Kuchowicz, B. 1978, GReGr, 9, 511

 [33] Garcia de Andrade, L. C. 2001, NCimB, 116, 1107

 [34] Pop{\l}awski, N. J. 2010, PhLB, 694, 181

 [35] Pop{\l}awski, N. J. 2011, PhLB, 701, 672

 [36] Pop{\l}awski, N. J. 2012, GReGr, 44, 1007

 [37] Pop{\l}awski, N. 2013b, AstRv, 8, 108

 [38] Kazanas, D. 1980, ApJL, 241, L59

 [39] Guth, A. H. 1981, PhRvD, 23, 347

 [40] Linde, A. 1982, PhLB, 108, 389

 [41] N. J. Poplawski, Phys. Lett. B 690 (2010) 73.

 [42] Poplawski, N. J. 2013, PhLB, 727, 575

 [43] Stachowiak, T., \& Szyd{\l}owski, M. 2007, PhLB, 646, 209

 [44] Novello, M., \& Perez Bergliaffa, S. E. 2008, PhR, 463, 127

 [45] Battefeld, D., \& Peter, P. 2015, PhR, 571, 1

 [46] Brandenberger, R., \& Peter, P. 2016, arXiv:1603.05834

 [47] Novikov, I. D. 1966, JETPL, 3, 142

 [48] Pathria, R. K. 1972, Natur, 240, 298

 [49] Frolov, V. P., Markov, M. A., \& Mukhanov, V. F. 1989, PhLB, 216, 272

 [50] Frolov, V. P., Markov, M. A., \& Mukhanov, V. F. 1990, PhRvD, 41, 
 
 383

 [51] Smolin, L. 1992, CQGra, 9, 173

 [52] Stuckey, W. M. 1994, AmJPh, 62, 788

 [53] Easson, D. A., \& Brandenberger, R. H. 2001, JHEP, 06, 024

 [54] Smoller, J., \& Temple, B. 2003, PNAS, 100, 11216

 [55] M. Forger, H. R\"{o}mer, Annals of Physics, 309 (2004) 306--389.

 [56] Y. N. Obukhov, V. N. Ponomarev and V. V. Zhytnikov, Gen. Rel. Grav. 
	
 21 1107 (1989)

 [57] Y. N. Obukhov, Int. J. Geom. Meth. Mod. Phys. 3 95 (2006).

 [58] S. M. Christensen, J. Phys. A: Math. Gen. 13 3001 (1980).

 [59] H. Shabani, and A. Hadi Ziaie, arXiv:1709.06512.

 [60] L. Zhang, S. Chen and J. Jing, 2018, Int. J. Mod Phys. D 27, 12, 
 
 1850110

 [61] Parker L. 1968, PhRvL, 21, 562

 [62] Parker L. 1969, PhRv, 183, 1057

 [63] Parker L. 1971, PhRvD, 3, 346

 [64] Parker L. 1971, PhRvD, 3, 2546

 [65] Zel'dovich Y. B. 1970, JETPL, 12, 307

 [66] Zel'dovich Y. B., Starobinskii, A. A. 1977, JETPL, 26, 252

 [67] Oztas A. M., Dil E., Smith M. L., 2018, MNRAS, 476, 451

 [68] Chen W., Wu, Y. S., 1990, Phys, Rev. D, 45, 4728

 [69] Melia F., 2007, MNRAS, 382, 1917

 [70] Vishwakarma R. G., 2001, Class. Quantum Grav., 18, 1159

\end{document}